# The Cluster Distribution as a Test of Dark Matter Models. II: The Dipole Structure

P. Tini Brunozzi[1,2], S. Borgani[2,3], M. Plionis[3,4], L. Moscardini[5] and P. Coles[6]

[1] *Dipartimento di Fisica, Università di Perugia, via A. Pascoli, I-06100 Perugia, Italy*

[2] *INFN Sezione di Perugia, c/o Dipartimento di Fisica dell'Università, via A. Pascoli, I-06100 Perugia, Italy*

[3] *SISSA – International School for Advanced Studies, via Beirut 2-4, I-34013 Trieste, Italy*

[4] *National Observatory of Athens, Lofos Nimfon, Thesio, 18110 Athens, Greece*

[5] *Dipartimento di Astronomia, Università di Padova, vicolo dell'Osservatorio 5, I-35122 Padova, Italy*

[6] *Astronomy Unit, School of Mathematical Sciences, Queen Mary & Westfield College, Mile End Road, London E1 4NS, UK*

20 June 1995

**ABSTRACT**
We study the dipole structure of an extended redshift sample of Abell/ACO clusters in order to infer the dynamical origin of the motion of the Local Group (LG). To further elucidate the constraints this motion places on dark matter models, we use numerical simulations based on an optimized version of the truncated Zel'dovich approximation which we have shown in previous work to provide a reliable representation of large-scale gravitational clustering. Taking advantage of their low computational cost, we run 20 realizations of each of six different dark matter models: four of these have a density parameter $\Omega_\circ = 1$, while the other two have $\Omega_\circ = 0.2$, one with and one without a cosmological constant term. For the Abell/ACO sample, we have evaluated the parameter $\beta = \Omega_\circ^{0.6}/b_{cl}$ (where $b_{cl}$ is the linear bias parameter for the clusters), which reaches its asymptotic value at $R_{conv} \simeq 160\,h^{-1}Mpc$. Convergence occurs at this scale whether calculations are performed in the LG or CMB frame, but the asymptotic value differs in these two cases: we find $\beta_{LG} = 0.15 \pm 0.04$ and $\beta_{CMB} = 0.25 \pm 0.06$, respectively. After identifying in the simulations those observers having local densities and peculiar velocities similar to those of the Local Group, we construct mock cluster samples around them reproducing the same observational biases, and apply to these mock samples the same method of analysis as we used for the Abell/ACO sample. We find that an alignment between the cluster dipole and observer velocity ('CMB' dipole) directions, such as that observed ($\Delta\theta \lesssim 20°$), should not be expected necessarily: much larger misalignment angles are often found in all models considered. This, together with the large observer-to-observer variance estimates of $\beta$, makes it difficult to place any firm constraints on cosmological models. This result suggests that the dipole analysis of the cluster distribution has a relevance that is *cosmographical*, rather than *cosmological*. Our results also demonstrate that the large amplitude and convergence depth of the observed cluster dipole cannot be taken as strong evidence either for or against a low–density Universe.

## 1 INTRODUCTION

The relationship between the peculiar acceleration and peculiar velocity of galaxies can, at least in principle, provide a means to determine the cosmological density parameter $\Omega_\circ$. According to the linear theory of gravitational instability, the contribution of density fluctuations within a spherical volume $V_R$ to the peculiar veloc-ity of an observer placed at the centre is given in terms of the density fluctuation field $\delta$ by

$$v_p = \frac{H_o f(\Omega_o)}{4\pi} \int_{V_R} d^3r \frac{r}{|r|^3} \delta(r) W(r), \quad (1)$$

where

$$\delta(r) = \frac{\rho(r) - \bar{\rho}}{\bar{\rho}}. \quad (2)$$

In eq.(1), $f(\Omega_o) \simeq \Omega_o^{0.6}$ to a good accuracy, virtually independently of the value of the cosmological constant (e.g. Peebles 1993); the window function $W$ filters out small-scale fluctuations and defines the "size" and "shape" of the observer through its functional dependence on $r$. As the size of the sphere is increased, the contribution to $v_p$ increases until it eventually saturates at scales where $\delta \ll 1$.

Since the only well-defined cosmic peculiar velocity is that of the Local Group of galaxies (LG hereafter), determined from the dipole pattern of the CMB radiation temperature (Kogut et al. 1993), the relation (1) is most naturally applied to the LG. By measuring the integral in eq.(1), we could obtain an estimate of $\Omega_o$. Although easy in principle, it is a hard task in practice, since we have no direct way to measure $\delta(r)$. The best we can hope to do is to use some class of cosmic objects as tracers of the underlying density field and hope that their point-like distribution, $n(r) = \sum_i \delta_D(r - r_i)$ is simply related to the underlying $\delta(r)$. In the previous expression $\delta_D$ stands for the Dirac delta-function. The assumption usually invoked is that relative fluctuations in the object number counts and matter density fluctuations are proportional to each other, at least within sufficiently large volumes, according to the linear biasing prescription (cf. Kaiser 1984; Bardeen et al. 1986):

$$\frac{\delta n(r)}{\bar{n}} = b \frac{\delta \rho(r)}{\bar{\rho}}, \quad (3)$$

where $b$ is what is usually called biasing parameter. If this is the case, then the comparison between velocity and acceleration provides a constraint on the parameter $\beta$, given by

$$\beta \equiv \frac{f(\Omega_o)}{b} \simeq \frac{\Omega_o^{0.6}}{b}. \quad (4)$$

It should be clear that in order to infer the $\beta$-parameter by relating the dipole estimated from an observational sample to the LG velocity, one must be sure that (*a*) the sample size is sufficiently large to contain all the fluctuations responsible for the LG motion, and (*b*) the CMB and the mass tracer dipoles are sufficiently well aligned to ensure that linear theory, as expressed by eq.(1), is applicable and that their estimates are not contaminated by the noise associated to the sparse sampling of the continuous density field.

The dipoles of various populations of extragalactic objects have been determined up to this date: optical galaxies (Lahav 1987; Plionis 1988; Lahav, Rowan-Robinson & Lynden-Bell 1988; Lynden-Bell, Lahav & Burstein 1989; Hudson 1993), IRAS galaxies (Meiksin & Davis 1986; Yahil, Walker & Rowan-Robinson 1986; Villumsen & Strauss 1987; Strauss & Davis 1988; Rowan-Robinson et al. 1990; Strauss et al. 1992; Plionis, Coles & Catelan 1993); X-ray active galactic nuclei (Miyaji & Boldt 1990); X-ray clusters (Lahav et al. 1989) and Abell clusters (Scaramella, Vettolani & Zamorani 1991; Plionis & Valdarnini 1991 [hereafter PV91]; Branchini & Plionis 1995). In all cases the dipole moment is reasonably well aligned with the CMB dipole suggesting that gravity is indeed responsible for the Local Group motion and that linear theory (eq. 1) can be applied to infer the value of $\beta$. However, these studies have provided conflicting estimates of $\beta$ which, in some cases, could be attributed to the different amount of biasing with respect to the background matter distribution (i.e. different $b$ values) but, in other cases, one would have to infer differences also in the value of $\Omega_o$ (for an overview, see Dekel 1994). In particular, IRAS galaxies have provided values $\beta_{IRAS} \approx 0.8 \pm 0.2$ implying $(\Omega_o, b_{IRAS}) \approx (1, 1.2)$, optical galaxies values somewhat smaller of $\beta_{opt} \approx 0.6 \pm 0.2$, implying $(\Omega_o, b_{opt}) \approx (1, 1.7)$ and clusters even smaller values $\beta_{cl} \approx 0.15 \pm 0.05$ implying either a quite large cluster biasing parameter ($b_{cl} \sim 7$) which is not supported by correlation function and power-spectrum studies (Peacock & Dodds 1994; Jing & Valdarnini 1993) or a value of $\Omega_o < 1$ (Plionis, Coles & Catelan 1993; Coles & Ellis 1994; see Branchini & Plionis 1995 for $\beta_{cl} \sim 0.2$ which is consistent with $\Omega_o \approx 1$).

In this work we plan to exploit the value of $\beta$ derived from the cluster distribution to address the following two questions:

(i) Is the use of linear theory to infer the value of $\beta_{cl}$ (and, possibly, of $\Omega_o$) from an observational setup similar to that of the real Abell/ACO cluster sample reliable?

(ii) Does the cluster dipole provide a strong constraint to discriminate between different Dark Matter (DM) models?

The paper is organized as follows. In Section 2 we explain our method of analysis. In Section 3 we describe the Abell/ACO cluster sample used and the results obtained in the analysis of the real cluster dipole. In Section 4 we present the simulation method applied to generate realistic cluster distributions and the 'observer' selection procedure. In Section 5 we discuss the results of our analysis of simulated cluster catalogues. Finally, in Section 6, we present our main conclusions. In the Appendix, some technical details on the spherical harmonics mask method are discussed.

## 2 THE DIPOLE ANALYSIS

Under the assumption of linear gravitational instability and linear biasing, eq.(1) relates the observer peculiar velocity to the distribution of clusters according to

$$v_p = \frac{H_o \beta}{4\pi \bar{n}} \sum_{i=1}^{N(R)} w_i \frac{r_i}{|r|_i^3} = \beta V_{cl} : \qquad (5)$$

$\beta = \Omega_o^{0.6}/b_{cl}$; $b_{cl}$ is the cluster biasing parameter; $N(R)$ is the total number of clusters within a distance $R$, $w_i$ are suitable weights which account for the different cluster masses and for selection biases in the observed cluster distribution (e.g. Galactic absorption, redshift selection, systematic biases, etc.). Estimating $V_{cl}$ from the observed cluster distribution and knowing $v_p$ from the CMB dipole amplitude, eq.(5) allows one to determine the parameter $\beta$.

It is useful to introduce the *monopole* and *dipole* of the cluster distribution evaluated at a scale $R$, as

$$M(R) = \frac{1}{4\pi} \sum_{i=1}^{N(R)} \frac{w_i}{r_i^2} \; ; \; D(R) = \frac{3}{4\pi} \sum_{i=1}^{N(R)} w_i \frac{r_i}{|r|_i^3}. \quad (6)$$

According to the above expressions, if all the mass (clusters) were concentrated at a single point, then $|D| = 3M$; for a completely isotropic distribution $D = 0$, while for a Poissonian distribution $|D| \sim 3M/\sqrt{N}$. In general the dipole amplitude is expected to increase with distance for as long as the sphere of radius $R$ has not encompassed the largest inhomogeneity in the cluster distribution. Its flattening is the signature that *isotropy* has been reached. The monopole grows as $M(R) \propto R$ for a uniform distribution and, therefore, if the convergence scale of the dipole $R_{conv}$ is within the sample size, then for scales $R \geq R_{conv}$ we have $M(R) = 4\pi \bar{n} R$. In what follows we define as the effective average cluster density the value derived from the monopole at the scale of the dipole convergence: $\bar{n} = M(R_{conv})/(4\pi R_{conv})$.

By inserting eqs.(6) into eq.(1), we get

$$V_{cl} = 3D(R) \frac{H_o R_{conv}}{M(R_{conv})} \qquad (7)$$

which will be used in the following as the estimator of $V_{cl}$. Note, however, that eqs.(6) can be directly applied only in the case in which the cluster sample covers the whole sky. This ideal situation is not satisfied in all the realistic cases: Galactic obscuration progressively degrades cluster detections at low Galactic latitudes and so affects the determination of both amplitude and direction of $D$. Different methods for correcting for this effect have been used by different authors. One approach is to fill the obscured portion of the sky uniformly (cf. Strauss & Davis 1988; Lahav, Rowan-Robinson & Lynden-Bell 1988). Another possibility is to "clone" the observed distribution near the zone of avoidance and extrapolate it across the Galactic plane (cf. Lynden-Bell, Lahav & Burstein 1989; Branchini & Plionis 1995). Yet an other approach is that based on the spherical harmonic reconstruction method. This method, which we will use in our present analysis, has been applied to the galaxy distribution by Yahil, Walker & Rowan-Robinson (1986), Lahav (1987), Plionis (1988, 1989) and, for the cluster distribution, by PV91. Details of this method can be found in the Appendix.

## 3 ANALYSIS OF THE ABELL/ACO CLUSTERS

### 3.1 The sample

The Abell/ACO sample (Abell 1958; Abell, Corwin & Olowin 1989) considered here is the same as that described in Borgani et al. (1995, Paper I) with additional new redshifts provided by the ESO Abell cluster survey (Katgert et al. 1995). It consists of all clusters having both $m_{10} < 17$, where $m_{10}$ is the magnitude of the tenth brightest cluster galaxy in the magnitude system corrected according to PV91, and an estimated distance $< 240 \; h^{-1}$ Mpc. There are in total 261 Abell clusters (dec $> -17°$), all having measured redshifts, and 201 ACO clusters, out of which 35 have $z$ estimated from the $m_{10}$-$z$ relation derived in PV91. Since we will compare the real data with simulations based both on flat and open cosmological models, with and without a cosmological constant, we convert redshifts into cluster distances using the general formula for the distance by apparent size:

$$r(z) = \frac{c}{H_o(1 - \Omega_o - \Omega_\Lambda)} \sinh\left(\frac{\int_0^z dz'/E(z')}{1 - \Omega_o - \Omega_\Lambda}\right), \quad (8)$$

where $c$ is the speed of light and

$$E(z) = \sqrt{\Omega_o(1+z)^3 + \Omega_\Lambda + (1 - \Omega_o - \Omega_\Lambda)(1+z)^2} \quad (9)$$

(e.g. Peebles 1993). Final results are actually quite insensitive to the choice of the cosmological parameters to be inserted in eq.(8) so, except where differently specified, we will from now on use distances based on $(\Omega_o, \Omega_\Lambda) = (0.4, 0)$, which are intermediate between the cases with $(\Omega_o, \Omega_\Lambda) = (1, 0)$ and $(\Omega_o, \Omega_\Lambda) = (0.2, 0.8)$ cases.

To a good approximation, the redshift selection function, $P(z)$, has the functional form (cf. Postman et al. 1989):

$$P(z) = \begin{cases} 1 & \text{if } z \leq z_c \\ A\,e^{-z/z_o} & \text{if } z > z_c \end{cases} \quad (10)$$

where $A = e^{z_c/z_o}$ and $z_c$ is the redshift out to which the space density of clusters remains constant (volume-limited regime). Using a best–fitting procedure, we obtain $z_c = 0.0786$, $z_o = 0.01$ and $z_c = 0.069$, $z_o = 0.009$ for Abell and ACO samples, respectively. Since correcting for this effect (by weighting clusters by the inverse of $P(z)$) could introduce large shot noise errors, because of the exponential decrease of the cluster density, we prefer to limit our analysis to $r_{max} = 240\,h^{-1}\,Mpc$. This is also consistent with the box size ($480\,h^{-1}\,Mpc$) of the simulations we shall discuss later (see below).

In Figure 1 we present the cluster number density estimated in equal *proper* volume shells and corrected for Galactic absorption, in both the Abell and ACO samples (with $|b| \geq 30°$). Within the volume-limited regime we have $n_{Abell} \simeq 1.5 \times 10^{-5}\,(h^{-1}\,Mpc)^{-3}$ and $n_{ACO} \simeq 2.3 \times 10^{-5}\,(h^{-1}\,Mpc)^{-3}$, respectively. These values correspond to average cluster separations of $\langle r_{Abell} \rangle \simeq 40\,h^{-1}\,Mpc$ and $\langle r_{ACO} \rangle \simeq 35\,h^{-1}\,Mpc$. The difference in densities of Abell and ACO clusters is partly intrinsic, due to the presence of the Shapley concentration in the ACO sample (Shapley 1930; Scaramella et al. 1989; Raychaudhury 1989), but probably mostly spurious, due to the higher sensitivity of the IIIa–J emulsion plates on which the ACO survey is based (for more details see Batuski et al. 1989; Scaramella et al. 1990; PV91). In order to account for the difference in the cluster number density we weight each Abell cluster according to a radial weighting function, $w(R) = n_{ACO}(R)/n_{Abell}(R)$, which accounts for a distance dependence of the number density difference. We have also verified that our final results are insensitive in weighting the Abell or ACO clusters with $w(R)$ or $1/w(R)$, respectively.

In order to account for cluster selection as a function of Galactic latitude, we model the Galactic absorption according to (see PV91):

$$\Theta(\vartheta) = \begin{cases} \exp(-\mathcal{A}\,|\cos\vartheta|^{-1}) & \text{if } |b| \geq |b|_{lim} \\ 0 & \text{if } |b| < |b|_{lim} \end{cases} \quad (11)$$

with $\vartheta = 90° - b$. We calculate the parameter $\mathcal{A}$ by requiring eq.(11) to fit the cluster data for $|b| \geq |b_{lim}|$ ($\equiv 13°$). This choice of $|b_{lim}|$ is a compromise between the need to have a large sky coverage and, in particular, to contain the Perseus cluster, which was shown in PV91 to have a significant effect on the cluster dipole, whilst simultaneously avoiding noise associated with regions of strong absorption. The best–fit values for $b \geq 13°$ (North Galactic Cap, NGC) and $b \leq -13°$ (South Galactic Cap, SGC) are $\mathcal{A}_N \simeq 0.75$ and $\mathcal{A}_S \simeq 0.63$, respectively. We use eq.(11) for the Galactic obscuration mask to work out the 'whole–sky' monopole and dipole coefficients (see Appendix).

### 3.2 Results

Dipole analyses of the Abell/ACO cluster distribution have been already presented by different authors (PV91; Scaramella, Vettolani & Zamorani 1991; Branchini & Plionis 1995). Here we briefly discuss our analysis of the data in order to make a homogeneous comparison with the simulation results. Note that here we do not weight clusters according to their mass, as done in most other similar studies, since we are interested in a consistent and meaningful comparison with results from our simulations, in which we cannot resolve meaningfully the individual cluster masses.

In Figure 2 we plot $V_{cl}$ evaluated according to eq.(7), after correcting the dipole and monopole estimates for Galactic absorption according to the spherical harmonic reconstruction method described in the Appendix. Upper and lower curves refer to estimates made in the LG and CMB frames, respectively. The estimate of the total dipole uncertainty includes Poissonian errors, the effects of uncertainties in the shape of $\Theta(\vartheta)$ and in the radial weighting function, which PV91 estimated to be $\sim 12\%$. For sake of clarity, we plot the total uncertainty only on the dipole value at the largest distance.

The overall shape of $V_{cl}(R)$ in both reference frames is quite similar: a steep rise up to $\sim 40\,h^{-1}\,Mpc$ reflecting the dynamical effect of the "Great Attractor" region, with a subsequent decrease due to the pull caused by the Perseus–Pisces region in the opposite direction at $\sim 50\,h^{-1}\,Mpc$. After that, $V_{cl}(R)$ remains rather flat up to $R \sim 100\,h^{-1}\,Mpc$ and, then, starts increasing again before reaching a point of apparent convergence, within the completeness limit of the sample, at about $R_{conv} \simeq 160\,h^{-1}\,Mpc$.

Despite the similarity of the LG and CMB dipole shapes, their asymptotic amplitudes turn out to be quite different, with $V_{cl}^{LG} = 3400 \pm 500\,km\,s^{-1}$ and $V_{cl}^{CMB} = 2000 \pm 300\,km\,s^{-1}$. This can be understood by noting that the distribution of clusters in redshift space differs from the real space (i.e. truly three–dimensional) distribution by the non–linear term:

$$cz = H_\circ |r| + [v(r) - v(0)] \cdot \frac{r}{|r|}, \quad (12)$$

**Table 1.** Abell/ACO cluster analysis. Columns 2 and 4: $\beta$ estimates in CMB and LG frames, after subtracting a Virgocentric infall of $200 \, km \, s^{-1}$. Columns 3 and 5: misalignment angle between the cluster dipole and the CMB dipole in the two frames. Results are obtained after correcting for the $\Theta$-mask.

| $R \, (h^{-1} Mpc)$ | $\beta_{CMB}$ | $\Delta\theta_{CMB}$ | $\beta_{LG}$ | $\Delta\theta_{LG}$ |
|---|---|---|---|---|
| 70  | $0.30 \pm 0.16$ | $24.3°$ | $0.19 \pm 0.11$ | $23.8°$ |
| 110 | $0.29 \pm 0.10$ | $19.9°$ | $0.17 \pm 0.06$ | $18.5°$ |
| 150 | $0.24 \pm 0.07$ | $22.3°$ | $0.15 \pm 0.05$ | $18.1°$ |
| 190 | $0.24 \pm 0.06$ | $20.1°$ | $0.15 \pm 0.04$ | $17.8°$ |
| 230 | $0.25 \pm 0.06$ | $22.1°$ | $0.15 \pm 0.04$ | $16.9°$ |

where $\mathbf{v}(\mathbf{r})$ is the peculiar velocity of a cluster at the position $\mathbf{r}$, in the observer frame, and $\mathbf{v}(0)$ is the observer peculiar velocity. In the LG frame (where $\mathbf{v}(0) = 620$ km/sec), structures placed in the direction of our motion appear at a redshift smaller than that measured in the CMB frame (where $\mathbf{v}(0) = 0$). These structures consequently enhance the apparent value of $V_{cl}$. One is tempted then to prefer to make calculations in the CMB frame. However, if nearby clusters, which dominate the dipole signal, participate together with the LG in a bulk motion [i.e. $v(r) \approx v(0)$] then a better frame in which to evaluate the cluster dipole would be the LG frame since in this case we have that $cz \approx H_0|r|$. These qualitative arguments serve only to indicate that one should consider the dipole estimated in the LG frame as an upper limit to the true 3-D dipole while that estimated in the CMB frame is a lower limit. In fact, a detailed 3-D reconstruction procedure used to recover the real-space cluster distribution shows that the value of the 3-D cluster dipole is roughly half way between the two above extremes (Branchini & Plionis 1995).

Using the above values of $V_{cl}$ and $v_{LG} = 620 \, km \, s^{-1}$, we find $\beta_{LG} = 0.18 \pm 0.04$ and $\beta_{CMB} = 0.33 \pm 0.06$. Note, however, that these values are overestimated with respect to the real ones because a significant part of the LG velocity is due to the pull of the nearby Abell-like Virgo cluster, which is not included in the Abell sample, due to its proximity. Therefore, a consistent comparison between $v_{LG}$ and $V_{cl}$ can be done only after subtracting the contribution of Virgo from the former. Allowing for a Virgocentric infall of $\sim 180 \, km \, s^{-1}$, we end up with $v_{LG} \simeq 500 \, km \, s^{-1}$, which brings the above estimates down to $\beta_{LG} \simeq 0.15$ and $\beta_{CMB} \simeq 0.25$.

In Table 1 we report the values of $\beta$ evaluated at different scales in the two frames, along with the misalignment angles $\Delta\theta$ with respect to the CMB dipole direction. Although the final $V_{cl}$ convergence depth occurs at $R_{conv} \simeq 160 \, h^{-1}$ Mpc, its direction is already quite well aligned with the CMB dipole at much smaller distances. Furthermore, the increased coherence, due to redshift space distortions and/or the existence of a bulk flow, produces a smaller misalignment angle in the LG frame, with $\Delta\theta_{LG} \simeq 18°$ while in the CMB frame $\Delta\theta_{CMB} \simeq 22°$.

The above values of $\beta$ can be compared with other estimates of $\beta$ based on the cluster distribution. PV91 found from a similar analysis $\beta_{LG} \simeq 0.19 \pm 0.03$. Using their method of reconstructing real-space cluster positions, Branchini & Plionis (1995) found $\beta \simeq 0.21$, with a large uncertainty ($\sim 15\%$), while Scaramella (1995) found $\beta_{LG} = 0.18 \pm 0.02$. By comparing the Abell/ACO cluster dipole amplitude with that of the IRAS QDOT galaxies, in the LG frame, Plionis (1995) estimated $b_{cl}/b_I \simeq 4$ for the ratio of the two biasing parameters while Peacock & Dodds (1994) suggest $b_{cl}/b_I \simeq 4.5$ by comparing the power-spectrum amplitude of clusters and IRAS galaxies. Together with their estimate of $\beta_I = 1.0 \pm 0.2$ from redshift space clustering distortions, this turns into $\beta_{cl} \simeq 0.22 \pm 0.05$ for the cluster distribution. All these estimates of $\beta_{cl}$ fall within the upper (CMB frame) and lower (LG frame) limits of our estimated $\beta$-values.

In the following Section, we will attempt to constrain different DM models by comparing our estimate of $\beta$ from the real data, with corresponding results from the cluster simulations.

## 4 THE SIMULATIONS

### 4.1 The method

The method we use to generate realistic cluster distributions is based on the Zel'dovich approximation (ZA, hereafter; Zel'dovich 1970; Shandarin & Zel'dovich 1989). The details of the simulation procedure are described in Borgani, Coles & Moscardini (1994) and Paper I, to which we refer the reader for further details. In the following we will sketch only the basic steps of our simulation procedure.

The ZA assumes that, at the time $t$, the final (Eulerian) positions $x(q, t)$ of self-gravitating particles are given by $x(q, t) = q + b(t) \nabla_q \psi(q)$. Here $\mathbf{q}$ are the initial (grid) particle positions, $b(t)$ is the growing mode of the evolution of linear density perturbations and $\psi(q)$ is the gravitational potential, which is related to the initial (linear) density fluctuation field, $\delta(q)$, through the Poisson equation $\nabla_q^2 \psi(q) = -\delta(q)/a$. Although gravity determines the particle displacements according to the ZA in a non-local fashion, no gravitational acceleration appears in the equation of motion. Therefore, particles move inertially along straight lines. The corresponding velocity,

$$v = \dot{b}(t)\,\nabla_q \psi(q)\,, \qquad (13)$$

is related to the initial fluctuation field according to linear theory. In order to reduce the effects of shell crossing, we filter out short–wavelength fluctuation modes according to $P(k) \to P(k)\exp(-k^2 R_f^2)$, where $R_f$ is chosen in such a way that the average number of streams at each point is 1.1 (see Paper I).

In this paper we mainly consider simulations of box size $L = 480\,h^{-1}\,Mpc$ which are larger than those presented in Paper I, which had $L = 320\,h^{-1}\,Mpc$. Since we consider the same number of particles and grid points (namely $128^3$), we have a lower resolution: the grid size is $3.75\,h^{-1}\,Mpc$. This is not a serious problem for the present analysis, since we do not need to know the cluster positions with great precision. We merely need to check how clusters trace the overall density field, and, then, the dynamical origin of the motion induced by gravity on suitably chosen observers. In fact, in Section 3 we have shown that, although the observed alignment between the CMB dipole and that traced by real clusters already occurs at relatively small distances ($\lesssim 50\,h^{-1}\,Mpc$), the dynamical contribution of clusters to the LG motion nevertheless saturates at scales considerably larger than this ($\sim 160\,h^{-1}\,Mpc$). We therefore expect that waves within a box of $320\,h^{-1}\,Mpc$ cannot provide by themselves a completely satisfactory description of the observational situation. Furthermore, in order to verify that the adopted box size is large enough, we will also refer in some cases to simulations based on a larger ($L = 640\,h^{-1}\,Mpc$) box size to corroborate our results.

Particles are initially placed on the grid points and are moved according to the ZA. The density and velocity fields on the grid are then reassigned through a TSC interpolation scheme (e.g. Hockney & Eastwood 1981) for the mass and the momentum carried by each particle. Clusters are then identified as the $N_{cl} = (L/\langle r_{cl}\rangle)^3$ highest local density maxima on the grid, where $\langle r_{cl}\rangle$ is the average cluster separation. In the following, we assume $\langle r_{cl}\rangle = 38\,h^{-1}\,Mpc$, so that we end up with a total of 2015 clusters in each realization. The above value of $\langle r_{cl}\rangle$ is half–way between that of the Abell and ACO samples ($\sim 40\,h^{-1}\,Mpc$ and $\sim 35\,h^{-1}\,Mpc$, respectively). In any case, we have verified that no significant changes in the statistical properties of the cluster dipole occurs when passing from $\langle r_{cl}\rangle = 38\,h^{-1}\,Mpc$ to either value of $\langle r_{cl}\rangle$.

We ran simulations for the six different models of the initial fluctuation spectrum. All the models, except the open one, are normalized to be consistent with the quadrupole of the CMB temperature anisotropy measured by COBE (Bennett et al. 1994). For each model we generated 20 random realizations. Four models have $\Omega_o = 1$ for the cosmic density parameter: the standard CDM model (SCDM), with $h = 0.5$ and $\sigma_8 = 1$ for the r.m.s. fluctuation amplitude within a top–hat sphere of $8\,h^{-1}\,Mpc$; the tilted CDM model (TCDM), with primordial spectral index $n = 0.7$, $h = 0.5$ and $\sigma_8 = 0.5$ (e.g. Adams et al. 1993; Moscardini et al. 1995); a low Hubble constant CDM model (LOWH), with $h = 0.3$ and $\sigma_8 = (1.6)^{-1}$ (Bartlett et al. 1994); the Cold+Hot DM model (CHDM), with $\Omega_{hot} = 0.3$ for the fractional density of the hot component, $h = 0.5$ and $\sigma_8 = (1.5)^{-1}$ (e.g. Klypin et al. 1993). Other two models have low density parameter ($\Omega_o = 0.2$) both with open (OCDM with $h = 1$ and $\sigma_8 = 1$) and with flat geometry ($\Lambda$CDM with $\Omega_\Lambda = 0.8$ for the cosmological constant term, $h = 1$ and $\sigma_8 = 1.3$). A problem with this latter model is that the r.m.s. value of $\sigma_8$ at the cluster scale ($\simeq 8\,h^{-1}\,Mpc$) exceeds unity. We do not therefore expect that the ZA will give a correct description of the cluster distribution in this case. Although this could be a problem for clustering analysis, which requires a precise determination of cluster positions, it is of secondary importance for the purpose of this work, where we just require that clusters roughly trace the large scale DM distribution. In a sense, while the determination of cluster correlations requires the knowledge of cluster positions within superclusters, questions concerning the amplitude and scale of convergence of the cluster dipole can be probably addressed just by knowing the positions of such superclusters.

In a recent paper, Gaztañaga, Croft & Dalton (1995) have attempted to cast doubt upon the reliability of the use of the ZA for cluster simulations like these. Their arguments are, however, based solely on a comparison between N–body and ZA approaches for the $\Lambda$CDM model, which we already know to be the most problematic for our method. In contrast, they fail to discuss the fact that, for the SCDM and CHDM models, they obtain similar results and arrive at the same conclusions as in our analysis based on the ZA (Plionis et al. 1995; Borgani et al. 1995). More details of tests we have made demonstrating the general reliability of our ZA cluster simulations for clustering studies are given in Borgani et al. (1995).

### 4.2 LG–observer identification

In order to perform a consistent comparison between simulations and real data, we need to reproduce the ob-servational set-up as closely as possible. To this end, we need to define 'observer' characteristics, such as their local density and velocity. In particular, a reliable definition of the 'observer' peculiar velocity is crucial, for at least two reasons: *(a)* the frequency for the occurrence of good alignment between $v_{obs}$ and $V_{cl}$ should tell us whether the small value of the observed $\Delta\theta$ (cf. Table 2) is quite typical or it is due to the presence of a highly coherent "attractor" in the nearby cluster distribution; *(b)* the comparison of the $|v_{obs}|$ with $|V_{cl}|$ could furnish an estimate of the parameter $\beta$.

We place observers at grid points and define both, their local density contrast and bulk velocity, averaged over a top-hat sphere of radius $R = 7.5\,h^{-1}Mpc$ centred on them. The local value of $\delta$ is estimated by convolving the Fourier transform of the fluctuation density field with the Fourier transform of the top-hat window,

$$W(kR) = \frac{3(\sin kR - kR \cos kR)}{(kR)^3}. \quad (14)$$

The definition of the observer's velocity is however less straightforward. We have considered two alternative ways of defining this quantity:

- The first is to use eq.(13) to define velocities for the particles and then to assign the corresponding momentum on the grid by means of the top-hat window function. The drawback of this procedure lies on the fact that the velocities of eq.(13) are in some sense not "synchronized" with the evolution of the density field; they refer to the initial field, $\delta(q)$, while the Zel'dovich evolution has moved structures away from their initial locations. Therefore, there is no reason to expect *a priori* any accurate alignment between the direction of the linear (initial) velocity vector, attached to the final particle position and the dipole estimated at that position by integrating over the ZA-evolved fluctuation field.
- The second criterion, which is in fact the one that we chose to adopt, is to define the velocity of a given 'observer' by means of eq.(5), where the dipole term is evaluated in a sphere over the whole distribution of the $128^3$ DM particles at a distance larger than $R = 7.5\,h^{-1}Mpc$ from the observer. Note that when the borders of the box are encountered we wrap-around, so that spherical symmetry is preserved. Here the value of the biasing parameter is set to unity by definition, while $\Omega_o$ is that of the model under consideration. This method of assignment assumes that, after ZA evolution, the gravitational potential and velocity fields are still connected by the linear relation $v \propto \nabla \psi$, but where $\psi$ is now the potential corresponding to the ZA evolved density field.

**Table 2.** Column 2: Number of observers satisfying the LG requirements. Column 3: r.m.s. observer velocity (in $km\,s^{-1}$). Columns 4 and 5: fraction of observers with misalignment $< 20°$ in LG and CMB frames, respectively.

| Model | $N_{LG}$ | $v_{rms}$ | $f(\Delta\theta_{LG})$ | $f(\Delta\theta_{CMB})$ |
|---|---|---|---|---|
| SCDM | 3019 | 592 | 0.38 | 0.21 |
| LOWH | 2366 | 461 | 0.32 | 0.20 |
| TCDM | 1347 | 384 | 0.28 | 0.17 |
| CHDM | 3192 | 593 | 0.39 | 0.22 |
| OCDM | 331 | 331 | 0.41 | 0.31 |
| $\Lambda$CDM | 974 | 399 | 0.44 | 0.32 |

Note that we estimate the local $\delta$ values and the peculiar velocity for 2000 observers chosen at random positions in each realization, so that we end up with a total of $4 \times 10^4$ observers for each DM model. We then select those 'observers' that have the same characteristics as the Local Group, the number of which can also be viewed as a first discriminative test for our DM models (c.f. Gorsky et al. 1989; Davis, Strauss & Yahil 1991; Strauss, Cen & Ostriker 1993; Tormen et al. 1993; Strauss et al. 1994; Moscardini et al. 1995). These characteristics are:

(1) Peculiar velocity $V_{LG} = 627 \pm 44\,km\,s^{-1}$ (error corresponding to $2\sigma$ uncertainties; Kogut et al. 1993) for a top-hat sphere of radius $R = 7.5\,h^{-1}Mpc$ centred on the observer;
(2) Density contrast within the same sphere in the range $-0.2 \leq \delta_{LG} \leq 1$.

A further requirement concerns the quietness of the local flow, which implies a small value for the local shear. Moscardini et al. (1995) argued that this kind of constraint is not very restrictive (see also Tormen et al. 1993). On the other hand, Schlegel et al. (1994) found that CHDM performs much better than SCDM in producing locally quite Hubble flows, which are similar to that observed around the Local Group. In any case, we do not expect any strong shear to be induced by the laminar flow description of the ZA, so we have decided not to impose it in our present analysis.

In Table 2 we report the fraction of observers which satisfy the LG requirements for all the models. Also reported are the r.m.s. values of velocities of these observers. In Figure 3 we plot the frequency distributions $P(\delta)$ and $P(v)$ for the local density and velocities of the selected observers. The errorbars in each bin repre-sent the $1\sigma$ scatter over the ensemble of 20 realizations, which is consequently an estimate of the cosmic variance. The solid curves, superimposed on $P(v)$, are the best–fitting Maxwellian distributions. We note that the fit is very good in all the cases, showing that the initial Gaussian statistics of the velocity field is preserved by the Zel'dovich dynamics. The shaded areas delineate the observational uncertainties on the $\delta_{obs}$ and $v_{obs}$ values. It is evident that a large fraction of observers for all the models has a local density within the observational interval. The probability, however, of reproducing the LG velocity depends quite strongly on the model. For instance, SCDM and CHDM reproduce the observational situation with a high probability. On the other hand, low–density models give lower velocities, with OCDM behaving even worse than $\Lambda$CDM, because of the lower normalization of $\sigma_8$. Consequently, adopting a substantially lower normalization for the OCDM model, $\sigma_8 \simeq 0.5$, according to some proposed COBE normalized open models (Kamionkowski & Spergel 1994; Gorski et al. 1995), the velocities produced would be even smaller and the probability of having a LG–like observer would be smaller still. As for TCDM and LOWH models, the lower velocities with respect to SCDM are also due to their lower $\sigma_8$ normalization.

## 5 RESULTS FROM SIMULATIONS

In this section we present a dipole analysis of the cluster simulations. After showing the importance of the requirement that CMB and cluster dipoles be aligned, we verify the ability of the spherical harmonic reconstruction method, which we applied on the real data samples, to recover information about the intrinsic (i.e. whole–sky) cluster distribution. To this end, we extract realistic data sets from the simulations by imposing the same observational biases as in the Abell/ACO sample, and analyze them on the same footing as in Section 3. Finally, we will compare results based on the real data and the simulations in order to put constraints on the DM models considered.

### 5.1 Cluster and CMB dipole alignment

A fundamental requirement to select realistic observers is the existence of a good alignment between the CMB and the cluster dipole vectors. Under the hypothesis that linear gravitational instability generates the large–scale motions and that clusters trace fairly the underlying matter density field, modulo the biasing factor $b$ defined in eq.(3), we should expect the two vectors to be aligned, as is the case for the observational samples (see Table 1). In order to verify whether such a good alignment is always to be expected, we estimate, for each observer the true cluster dipole vector corresponding to the whole cluster distribution within $240\, h^{-1}\, Mpc$ (i.e. without including observational selection effects) and the misalignment angle $\Delta\theta$ with respect to $v_p$.

In Figure 4 we plot the $\beta_{LG}$ values estimated by using eq.(5) for 5000 $\Lambda$CDM and CDHM observers as a function of the corresponding $\Delta\theta$ misalignment angles between their CMB and cluster dipoles. The dipole amplitude is measured for all the observers at the scale $R_{max} = 230\, h^{-1}\, Mpc$. Also plotted is the corresponding cumulative $\Delta\theta$ distributions, both in the LG (solid curves) and in the CMB (dashed curves) frames. At least 95% of observers have $\Delta\theta < 90°$, thus showing that the cluster distribution does indeed reflect the dynamical origin of the observers' motion. We can summarize the main conclusions of the $\Delta\theta$ analysis by noting that:

- Dipoles in the LG frame have systematically a better alignment than those in the CMB frame, in agreement with the qualitative explanation given in Section 3.2 and with what we have found in the observational case.
- The difference between LG and CMB frames is smaller for $\Lambda$CDM than for CHDM, as expected from eq.(12) since the observers have smaller peculiar velocities in the former model.
- Due probably to the sparseness of the cluster distribution, most simulation observers measure a much larger value of $\Delta\theta$ than the values found for the Abell/ACO clusters and there is therefore no *a priori* reason to expect an alignment as accurate as that seen in the real data. This shows that the LG motion is probably generated by a large, very coherent structure, which is well sampled by the observed cluster distribution.

In order to reproduce the observational situation in the following analysis, we will select as LG candidates only those observers that measure $\Delta\theta \leq 20°$ in both CMB and LG frames. This is quite a conservative restriction since, as shown in Figure 4, no substantial changes in the $\beta$ estimates should occur for $\Delta\theta \lesssim 40°$. In fact, we have verified that taking this larger limit for $\Delta\theta$ does not change results in either the LG or CMB frame. In Table 2 we list the percentage of observers passing only this test for the different models.

Although no stringent constraints can be posed on the models from the statistics of $\Delta\theta$, it turns out that low–density models produce more relatively better "aligned" observers, especially in the CMB frame.

This could well be due to the fact that such models, which have more large–scale power, produce more coherent and extended structures in the cluster distribution. In the cases with $\Omega_o = 1$ observers, the less accurate alignment in the CMB frame is compensated in the LG frame by their higher velocities with respect to the $\Omega_o = 0.2$ models.

### 5.2 Estimates of the $\beta$–parameter

*5.2.1 Whole box analysis*

In Figure 5 we plot the frequency distribution of $\beta$ values measured by all the observers passing the $\Delta\theta$ test, for both $\Lambda$CDM and CHDM. In each panel, we report results of the analysis done in real space (solid line), in the CMB frame (dotted line) and in the LG frame (dashed line). We conclude that:

- The $\beta$ estimates in the CMB frame are always very close to the "true" ones of real space, which is evidence that most LG–like observers in our simulations are not embedded in any coherent bulk flow (see discussion in Section 3.2)*.
- The $\beta$ estimates in the LG frame are systematically larger than those in the CMB frame, which is to be expected according to eq.(12).
- The $\Lambda$CDM model produces systematically lower $\beta$ values than CHDM. The $P(\beta)$ distributions are, however, too broad to allow strong discrimination even between these two cases.

In order to verify whether the presence of fluctuation modes on scales larger than the box size can alter our results, we also analyzed simulations in a box of side $L = 640\,h^{-1}Mpc$. We verify that increasing the box size does not alter the $P(\beta)$ distributions significantly, while we also obtain very similar values of $\bar{\beta}$. For example, in the $\Lambda$CDM model, we get $\bar{\beta}_{CMB} = 0.16 \pm 0.06$ for the $L = 480\,h^{-1}Mpc$ box and $\bar{\beta}_{CMB} = 0.16 \pm 0.08$ for the $L = 640\,h^{-1}Mpc$ box. Therefore, wavelength modes on scales comparable to or exceeding the box size used, act in the same way on the value of $V_{cl}$ and on the observer velocity $v_p$, so as to make their ratio (i.e. the $\beta$ parameter) almost independent of the box size.

---

* In future we plan to investigate whether the existence or not of 'coherent motions' in our simulations could discriminate between the different DM models.

**Table 3.** Columns 2 and 4: Average $\bar{\beta}$ value for the mock samples, after correction for the $\Theta(\vartheta)$ mask in the CMB and LG frames, respectively. Columns 3 and 5: Number of LG–like observers measuring a $\beta$ value in the observational range after correcting for the $\Theta(\vartheta)$ mask for CMB and LG analyses, respectively.

| Model | CMB frame | | LG frame | |
|---|---|---|---|---|
| | Mask | $N_{real}$ | Mask | $N_{real}$ |
| SCDM | $0.23 \pm 0.14$ | 181 | $0.15 \pm 0.11$ | 492 |
| LOWH | $0.20 \pm 0.12$ | 107 | $0.13 \pm 0.15$ | 195 |
| TCDM | $0.19 \pm 0.11$ | 170 | $0.12 \pm 0.09$ | 340 |
| CHDM | $0.22 \pm 0.13$ | 238 | $0.14 \pm 0.10$ | 435 |
| OCDM | $0.16 \pm 0.07$ | 45 | $0.09 \pm 0.06$ | 75 |
| $\Lambda$CDM | $0.17 \pm 0.09$ | 122 | $0.11 \pm 0.07$ | 216 |

*5.2.2 Analysis after including realistic selection effects*

Observational biases, like Galactic obscuration and redshift selection, are expected to affect dipole determinations, especially when they are not isotropically distributed around the sky. Furthermore, the redshift selection function could significantly affect the dipole estimation especially if the redshift completeness limit is lower than the true dipole convergence depth. However, since we limit our analysis to $R_{max} = 230\,h^{-1}Mpc$, a depth at which the cluster dipoles have converged to their final values and at which the observed Abell/ACO cluster sample is quite complete, its effect should be negligible.

In order to account properly for these effects, we extract cluster samples with the same $\Theta(\vartheta)$ and $P(z)$ selection functions as the Abell/ACO sample [see eqs. (11) and (10), respectively], for each LG–like observer passing the misalignment test. Since the effect of the $\Theta(\vartheta)$ mask could depend on the Galactic latitude of the CMB dipole direction, we choose the observer coordinate system so that its $v_p$ has the same direction ($l = 277°, b = 30°$) as the observed CMB dipole. For each of these observers we construct mock catalogues by performing 100 Monte–Carlo realizations of the selection functions, in order to minimize the noise due to accidental cluster selection in regions of strong Galactic obscuration. The resulting dipole amplitude of a given observer is taken to be the average over all those Monte Carlo realizations for which the misalignment angle still satisfies the constraint $\Delta\theta \leq 20°$ after the mask correction has been applied (as in the real cluster case).

In Figure 6 we plot the frequency distribution of the $\beta$ values estimated in the CMB frame, for all the LG–like

ΛCDM and CHDM observers, passing the three tests for $v_{LG}$, $\delta_{LG}$ and $\Delta\theta$. Intrinsic dipole $\beta$–values (solid histograms) are compared to those before mask correction (left panels) and after mask correction (right panels). The effect of introducing observational biases is that of increasing the noise, i.e. the width of the distribution. As far as the mask correction procedure is concerned we can conclude from this study that:

- *Dipole alignment*: The number of well–aligned dipoles after correcting for Galactic absorption is $\sim 20-30\%$ larger than that before the corrections; this result shows that the correction procedure is effective in recovering on average the cluster dipole direction.
- *Dipole amplitude*: The intrinsic $P(\beta)$ distribution is recovered quite well after the mask correction, confirming again the reliability of our procedure.

In Figure 6 we also plot, as the dashed band, the $1\sigma$ uncertainty in the the Abell/ACO $\beta$ estimate (c.f. Table 1). We note that the observational $\beta$–range is more typical in a CHDM universe, while it falls on the high–$\beta$ tail of the $P(\beta)$ distribution for the ΛCDM model.

In Table 3 we report the average values, $\bar{\beta}$, measured by LG–like observers for the realistic cluster samples, in both the LG and CMB frames, for all the models (after correcting for Galactic absorbtion). Although not reported here, the whole–box $\bar{\beta}$ values are very similar to those of Table 3, as indicated by the similarity between the corresponding $P(\beta)$ distributions (see Figure 6). Also reported is the number $N_{real}$ of such observers, which besides satisfying the LG and alignment tests, measure a $\beta$ value consistent with the real one.

Despite the fact that differences exist between the different models, all of them produce an appreciable number of such observers. This means that:

- It is quite difficult to discriminate at a high confidence level between different initial power–spectra, which is supported also by the large uncertainties in the $\bar{\beta}$ estimate due to the large spread of the $P(\beta)$ distributions.
- The Abell/ACO cluster dipole cannot be taken as strong evidence for a low–$\Omega_\circ$ Universe, as has been suggested by Scaramella et al. (1991) and PV91. Instead, the indications are, if anything, that some $\Omega_\circ = 1$ models are marginally preferred to low–density ($\Omega_\circ = 0.2$) models.

It is interesting to note that the Abell/ACO data show a difference between $\beta_{LG}$ and $\beta_{CMB}$ which is larger than the typical difference found in the simulated data sets. This further indicates that the observational situation is probably characterized by the presence of a large coherent structure, which causes a bulk motion of the Local Universe of an amplitude similar to the velocity of the Local Group (cf. PV91; Plionis, Coles & Catelan 1993; Branchini & Plionis 1995) thus generating, in the manner described by eq.(12), relatively larger differences between LG– and CMB–estimated cluster distances than those found in the simulations.

## 5.3 Estimating the cosmological density parameter

We can now investigate whether the estimated values of $\beta$ turn into a reliable estimate of the density parameter $\Omega_\circ$. In order to pass from $\beta$ to $\Omega_\circ$ one must know, in principle, the *local* biasing parameter, i.e. the proportionality constant which relates cluster and DM distributions *point by point*. What is usually estimated, however, is the global biasing factor $b_{cl}$, evaluated in a statistical sense from global correlation properties of the cluster distribution. Nonetheless, the non–linear selection of clusters as peaks of the density field means that the correct biasing parameter is actually a function of the position.

In the simulations we estimate $b_{cl}$ statistically, as the ratio between the r.m.s. fluctuation amplitude for cluster and DM distributions at different scales,

$$b_{cl}(R) = \frac{\sigma_{cl}(R)}{\sigma_{DM}(R)}. \qquad (15)$$

We estimate the r.m.s. fluctuations in the DM distribution as

$$\sigma_{DM}^2(R) = \frac{1}{2\pi^2} \int_0^\infty dk\, k^2 P(k)\, W^2(kR), \qquad (16)$$

where $W(kR)$ is the Fourier transform of the top–hat window function, as given by eq.(14), and $P(k)$ the linear power–spectrum.

Since the dipole analysis is based on counts within spheres of different radii, we estimate $\sigma_{cl}(R)$ through counts of clusters within top–hat spheres of radius $R$. Furthermore, as in the cluster dipole analysis, we do not correct the variance of the cluster distribution for Poissonian shot–noise. The analysis is done by randomly placing 5000 spheres within each simulation box, with radii ranging from $50\,h^{-1}Mpc$ and $120\,h^{-1}Mpc$. Although we need to estimate the value of $b_{cl}$ at the largest considered scale, $R_{max} = 230\,h^{-1}Mpc$, it turns out that, at this scale, the variance analysis is dominated by the finite box size by the consequent effect of periodic boundary conditions. However, consistently with what we found in Paper I, we have verified that $b_{cl}$ is almost independent of scale in the range we have explored. We therefore feel justified in extrapolating the value of $b_{cl}$ determined on smaller scales up to $R_{max} = 230\,h^{-1}Mpc$. The $b_{cl}$ values for all six of our models are reported in Table 4, along with the corresponding "cosmic" r.m.s. uncertainties, estimated from the 20 realizations of each model. The central value is used to convert the $\beta$ estimate of each observer into the corresponding $\Omega_o$ estimate.

Table 4. Column 2: Cluster biasing parameters and relative "cosmic" r.m.s. uncertainties. Column 3: $\Omega_o$ estimates for LG–like observers in the CMB frame for whole–box cluster samples. Column 4: The same as in column 3, but after introducing the $\Theta(\vartheta)$ mask and correcting for it. Errors come from uncertainties in the $\beta$ estimate (cf. Table 3).

| Model | $b_{cl}$ | $\langle\Omega_o\rangle$(whole sky) | $\langle\Omega_o\rangle$(masked) |
|---|---|---|---|
| SCDM | $3.74 \pm 0.17$ | $1.0^{+1.3}_{-0.8}$ | $0.8^{+1.1}_{-0.6}$ |
| LOWH | $5.09 \pm 0.22$ | $1.1^{+1.2}_{-0.9}$ | $1.0^{+1.3}_{-0.8}$ |
| TCDM | $7.43 \pm 0.39$ | $1.9^{+2.1}_{-1.4}$ | $1.8^{+2.0}_{-1.4}$ |
| CHDM | $4.11 \pm 0.16$ | $1.0^{+1.3}_{-0.8}$ | $0.9^{+1.5}_{-0.7}$ |
| OCDM | $2.96 \pm 0.14$ | $0.23^{+0.15}_{-0.03}$ | $0.29^{+0.24}_{-0.18}$ |
| $\Lambda$CDM | $2.36 \pm 0.11$ | $0.20^{+0.13}_{-0.11}$ | $0.22^{+0.22}_{-0.16}$ |

In Figure 7 we plot the distribution of $\Omega_o$ estimated in the CMB frame for the CHDM and $\Lambda$CDM LG–like observers. For both models, the $P(\Omega_o)$ distribution turns out to be rather broad and a non–negligible probability exists to measure values of $\Omega_o$ substantially different than the correct ones. In Table 4 we report, for all the models, the average $\Omega_o$ values measured by LG–like observers in the CMB frame. The two columns refer to estimates from the whole–box cluster distributions and from the mock cluster samples. Although from Figure 7 we note that the $P(\Omega_o)$ peaks at $\Omega_o$ values smaller than the correct ones, the average value always agrees well with the "intrinsic" $\Omega_o$. As for the comparison between low–density and high–density models, it turns out that there is a non–negligible probability of measuring $\Omega_o$ as low as 0.2 in a $\Omega_o = 1$ universe. Vice versa, if $\Omega_o = 0.2$ it is almost impossible to measure, by chance, $\Omega_o = 1$. On the other hand, values as large as 0.4 are commonplace is both of these models.

## 6 DISCUSSION AND CONCLUSIONS

We have compared the 3D dipole structure of an extended sample of Abell/ACO clusters, supplemented with the new redshifts of Katgert et al. (1995), to that generated by numerical simulations of the cluster distribution, based on six different DM models. The simulation technique, based on the Zel'dovich approximation, is so computationally cheap that it allows us to run a set of 20 realizations for each model. We present our main conclusions in the following subsections:

### 6.1 Real Cluster Dipole Results

We have estimated the Abell/ACO cluster dipole accounting for the effects of Galactic obscuration by applying the spherical harmonic method (see Appendix). Our results (see Table 1 and Figure 2) are in general agreement with those coming from other studies, based however on fewer measured cluster redshifts (PV91; Scaramella et al. 1991; Branchini & Plionis 1995) and can be summarized as follows.

(a) The dipole convergence occurs at a scale, $R_{conv} \simeq 160\,h^{-1}Mpc$, substantially larger than the convergence scale indicated by shallower galaxy samples (e.g. Strauss & Davis 1988; Lynden–Bell et al. 1989; Rowan–Robinson et al. 1990), but still well within the completeness limit ($\simeq 240\,h^{-1}Mpc$) of the cluster sample.
(b) Despite the large value of $R_{conv}$, the direction of the cluster dipole is already very well aligned ($\Delta\theta \lesssim 20°$) with the CMB dipole (Kogut et al. 1993) at much smaller scales ($\simeq 50\,h^{-1}Mpc$). This suggests the presence of an extended coherent puller generating the motion of the Local Group.
(c) The final value of the dipole amplitude, estimated both in LG and CMB frames, turns into values of the $\beta$–parameter of $\beta_{LG} = 0.15 \pm 0.04$ and $\beta_{CMB} = 0.25 \pm 0.06$. The difference between these two values is due to the different distances at which clusters are placed, according to eq.(12), along the direction of our motion.

### 6.2 Simulation Cluster Dipole Results

We took particular care to reproduce the actual observational setup as closely as possible in our simulated cluster distributions. This was realized in two steps:

- We identify those 'observers' which have local (i.e. within a sphere of $7.5\,h^{-1}Mpc$ radius) values for the density fluctuation $\delta_{obs}$ and peculiar velocity $v_{obs}$ consistent with those of the LG; i.e. $-0.2 \leq \delta_{LG} \leq 1$ and $v_{LG} = 627 \pm 44$ km s$^{-1}$. As a further criterion, we also require that the misalignment angle $\Delta\theta$ between the direction of the whole–sky cluster dipole and the observer velocity is $\Delta\theta \leq 20°$.
- For each of these LG–like observers we generate mock cluster samples having the same character-

istics (i.e. number of clusters, Galactic latitude and redshift selection functions) as the real sample. Also, for each observer the coordinate system is suitably rotated so that the vector $\mathbf{v}_{obs}$ points in the same direction as the observed CMB dipole.

The main results concerning the simulation analysis and their comparison with real data can be summarized as follows.

(a) The LG–like observers found in each model (Table 2 and Figure 3) are quite typical in $\Omega_o = 1$ models with a moderate $\sigma_8$ normalization, like SCDM and CHDM, but are much less likely in the $\Omega_o = 0.2$ models (OCDM and $\Lambda$CDM) and in a $\Omega_o = 1$ models with a low normalization ($\sigma_8 \simeq 0.5$), like TCDM.

(b) Regarding the frequency of occurrence of small misalignment angles between the cluster dipoles and the observer velocities, we find that small $\Delta\theta$ values are not to be expected in most cases. In fact, $\Delta\theta \leq 20°$ is measured by 30–40% of the observers in LG frame and by 20–30% of them in CMB frame (Table 2 and Figure 4).

(c) The $\beta$ estimates for real–space and CMB frame are quite similar, while analyses in the LG frame give systematically lower $\beta$'s. Although we estimate $\beta_{LG} < \beta_{CMB}$ also for real Abell/ACO clusters, Branchini & Plionis (1995) found that their real–space estimate of $\beta$ is about half way between $\beta_{LG}$ and $\beta_{CMB}$. This indicates that the observational situation is characterized by a coherent motion of the Local Group, which also involves the nearest clusters (see also Table 7 of PV91 and Plionis, Coles & Catelan 1993). In fact, if this is the case, then eq.(12) would imply that the cluster redshifts in the CMB frame are artificially overestimated.

(d) Although the $P(\beta)$ distributions are significantly different for different models, especially when comparing low–density and high–density models (see Figure 6), both the width of $P(\beta)$ and the rather large uncertainties of the observational $\beta$ value make it difficult to constrain any model. We point out, however, that SCDM and CHDM models are marginally preferred over the others, in the sense that they have the largest number of observers measuring a $\beta$ value consistent with the real one, while LOWH, OCDM and $\Lambda$CDM are less favoured (see Table 3).

(e) By directly estimating the cluster biasing parameter $b_{cl}$ in our simulations, we can determine the value of $\Omega_o$ that would be estimated by each observer. It turns out that single observers can measure $\Omega_o$ values very different from the true value. The correct value is, however, nicely recovered on average (see Table 4 and Figure 7).

### 6.3 Final Comments

Based on these results, we can safely state that the answers to the two questions raised in the Introduction are both negative. That is:

- Due (probably) to the sparse sampling of the underlying density field and the effect of cosmic variance, the estimated value of $\beta$ from Abell/ACO type cluster distributions can be very different from the true value. It therefore seems that we cannot infer the value of $\Omega_o$ with precision.
- No strong constraints can be posed on DM models based solely on the study of the dipole structure (its large amplitude and convergence depth) of the cluster distribution.

A further possibility to discriminate between low– and high–density cosmological models is offered by study of the large–scale cluster peculiar velocity field (see, e.g., Bahcall, Gramann & Cen 1994; Cen, Bahcall & Gramann 1994; Croft & Efstathiou 1994). The recent results of Lauer & Postman (1994) on the presence of a large amplitude cluster bulk flow, within $\lesssim 15000$ km/sec, presents a strong challenge for current cosmological models (Strauss et al. 1994; Feldman & Watkins 1994; Jaffe & Kaiser 1994). However, Branchini & Plionis (1995) and Branchini, Plionis & Sciama (1995), using a reconstruction method and linear gravitational instability theory to obtain the 3D cluster positions and peculiar velocities, found that the Lauer & Postman cluster sample exhibits a bulk flow of only $\sim 200$ km/sec, in much better agreement with most cosmological models.

We plan in the near future to take advantage of our ZA–based cluster simulations to test several DM spectra against observations of large–scale cluster motions, with particular emphasis on distinguishing between high– and low–density models.

**Acknowledgments.**

This work has been partly supported by funds originating from the *EC* Human Capital and Mobility Network (Contract Number CHRX–CT93–0129). PC is a PPARC Advanced Research Fellow. PTB and LM thank Italian MURST for partial financial support. We all thank the ESO *Nearby Abell Cluster Survey* team and in particular E. Escalera and A. Biviano for providing us with their cluster redshifts prior to publication.

## APPENDIX A1: SPHERICAL HARMONICS MASK METHOD

The method we use to treat the Galactic absorption effects is based on expanding the sky surface density field $\sigma(\vartheta, \varphi)$ in spherical harmonics according to

$$\sigma(\vartheta, \varphi) = \sum_{l=0}^{\infty} [a_l^0 P_l(\cos\vartheta) \qquad (A1)$$
$$+ \sum_{m=1}^{l} P_l^m(\cos\vartheta)(a_l^m \cos m\varphi + b_l^m \sin m\varphi),$$

where $0 \leq \vartheta < \pi$, $0 \leq \varphi < 2\pi$ and $P_l^m(\cos\vartheta)$ are the Legendre polinomials. Accordingly, the monopole is defined as $M = a_0^0$, while $|D| = [(a_0^1)^2 + (a_1^1)^2 + (b_1^1)^2]^{1/2}$ for the dipole amplitude.

If $\Sigma(\vartheta, \varphi)$ is the observed surface density field and $\sigma(\vartheta, \varphi)$ is the intrinsic one, the two are related by the mask function $\Theta(\vartheta, \varphi)$, which defines the missing parts of the survey, according to

$$\Sigma(\vartheta, \varphi) = \Theta(\vartheta, \varphi) \sigma(\vartheta, \varphi). \qquad (A2)$$

If we are interested in recovering the dipole ($l = 1$) components of $\Sigma(\vartheta, \varphi)$, correction terms in $\Theta(\vartheta, \varphi)$ should at least involve the quadrupole ($l = 2$) terms. We show below the detailed procedure to calculate the $(a_l^m, b_l^m)$ coefficients for the spherical harmonic expansion of $\sigma(\vartheta, \varphi)$ in terms of the observed $(A_l^m, B_l^m)$ coefficients, for a given mask model $\Theta(\vartheta, \varphi)$. Since in the case of Galactic absorption there is no $\varphi$ dependence of the mask we have that $\Theta(\vartheta, \varphi) = \Theta(\vartheta)$.

Starting from the expansion of $\Sigma(\vartheta, \varphi)$ up to the quadrupole order and allowing for the orthogonality relation of the Legendre polynomials,

$$\int_{-1}^{1} P_l^m(\mu) P_{l'}^m(\mu) d\mu = \frac{2\delta_{ll'}}{(2l+1)} \frac{(l+m)!}{(l-m)!} \qquad (A3)$$

($m \leq l, l'$), we can express the observed coefficients in terms of the intrinsic ones as

$$A_l^m = \frac{(2l+1)}{2\pi} \frac{(l-m)!}{(l+m)!} \qquad (A4)$$
$$\int_0^{2\pi} \int_0^{\pi} \sigma(\vartheta, \varphi) \Theta(\vartheta) P_l^m(\cos\vartheta) \cos m\varphi \sin\vartheta d\vartheta d\varphi$$

$$B_l^m = \frac{(2l+1)}{2\pi} \frac{(l-m)!}{(l+m)!} \qquad (A5)$$
$$\int_0^{2\pi} \int_0^{\pi} \sigma(\vartheta, \varphi) \Theta(\vartheta) P_l^m(\cos\vartheta) \sin m\varphi \sin\vartheta d\vartheta d\varphi.$$

We can express the mask $\Theta(\vartheta)$ as a function of the different Galactic absorption coefficients (for the NGC and SGC) by defining the following quantity:

$$H_n(\mathcal{A}) = \int_0^{\chi} \cos^{n-2}\vartheta \sin\vartheta \exp\left(\frac{-\mathcal{A}}{|\cos\vartheta|}\right) d\vartheta, \qquad (A6)$$

where $\chi = \pi/2 - |b|_{lim}$ and we introduce

$$H_n^+ = H_n(\mathcal{A}_N) + H_n(\mathcal{A}_S)$$
$$H_n^- = H_n(\mathcal{A}_N) - H_n(\mathcal{A}_S), \qquad (A7)$$

where $\mathcal{A}_N$ and $\mathcal{A}_S$ are the values of the Galactic extinction amplitude in the North and South Galactic Caps, respectively. Therefore, substituting the expansion (A1) of $\sigma(\vartheta, \varphi)$ and the above expression for the mask (11), eqs.(A4) and (A5) give a linear system of 9 equations with the 9 unknown quantities $a_l^m, b_l^m$ ($l \leq 2, m \leq l$) which can be written as:

$$A_0^0 = \frac{a_0^0}{2} H_2^+ + \frac{a_1^0}{2} H_3^- + \frac{a_2^0}{4} \left(3H_4^+ - H_2^+\right)$$

$$A_1^0 = \frac{3a_0^0}{2} H_3^- + \frac{3a_1^0}{2} H_4^+ + \frac{3a_2^0}{4} \left(3H_5^- - H_3^-\right)$$

$$A_1^1 = \frac{3a_1^1}{4} \left(H_2^+ - H_4^+\right) + \frac{9a_2^1}{4} \left(H_3^- - H_5^-\right)$$

$$B_1^1 = \frac{3b_1^1}{4} \left(H_2^+ - H_4^+\right) + \frac{9b_2^1}{4} \left(H_3^- - H_5^-\right)$$

$$A_2^0 = \frac{5a_0^0}{4} \left(3H_4^+ - H_2^+\right) + \frac{5a_1^0}{4} \left(3H_5^- - H_3^-\right) +$$
$$+ \frac{5a_2^0}{8} \left(9H_6^+ - 6H_4^+ + H_2^+\right)$$

$$A_2^1 = \frac{5a_1^1}{4} \left(H_3^- - H_5^-\right) + \frac{15a_2^1}{4} \left(H_4^+ - H_6^+\right)$$

$$B_2^1 = \frac{5b_1^1}{4} \left(H_3^- - H_5^-\right) + \frac{15b_2^1}{4} \left(H_4^+ - H_6^+\right)$$

$$A_2^2 = \frac{15a_2^2}{16} \left(H_2^+ - 2H_4^+ + H_6^+\right)$$

$$B_2^2 = \frac{15b_2^2}{16} \left(H_2^+ - 2H_4^+ + H_6^+\right).$$

By inverting the above system of equations we can work out the corrected monopole, dipole and quadrupole coefficients in terms of the observed ones.

**Figure Captions**

**Figure 1.** The Abell (solid line) and ACO (dashed line) cluster density for $|b| \geq 30°$, as a function of redshift. Poissonian errorbars are shown only for the Abell sample.

**Figure 2.** The dipole amplitude $V_{cl}$ for the Abell/ACO cluster sample, estimated both in the CMB (open circles) and in the LG (filled circles) frames after the spherical harmonic correction for the effects of Galactic extinction. The errorbars at the largest considered scale include both Poissonian errors and the effects of the uncertainties on the shape of the selection functions (see text).

**Figure 3.** The frequency distribution for the local density contrast $\delta_{obs}$ and the peculiar velocity $v_{obs}$ within a top-hat sphere of $7.5\,h^{-1}Mpc$ radius, for the 40,000 observers selected in each model. Errorbars correspond to the scatter between the 20 realizations of each model, while the solid curves for $P(v)$ are the Maxwellian best–fit. Shaded areas correspond to the observational values for local density $\delta_{obs}$ and velocity $v_{obs}$ of the Local Group (see text).

**Figure 4.** Scatter plots of the $\beta$–parameters, measured in LG frame by 5000 observers in $\Lambda$CDM (left panel) and CHDM (right panel) simulations, as a function of the misalignment angle $\Delta\theta$ between the directions of CMB and cluster dipole. Results refer to the whole–box simulated cluster distributions seen by each observer. Also shown in the top panels is the probability of having an alignment better that a given $\Delta\theta$, both in LG (solid curves) and in CMB (dashed curves) frames.

**Figure 5.** The $\beta$–parameter frequency distributions as measured by LG–like observers, identified in $\Lambda$CDM (left panel) and CHDM (right panel) simulations. The results come from the whole–box cluster distributions. Solid, dotted and dashed histograms refer to analyses realized in real space, CMB frame and LG frame, respectively.

**Figure 6.** The frequency distributions of $\beta$ values measured in the CMB frame by LG–like observers identified in CHDM (upper panels) and in $\Lambda$CDM (lower panels) simulations. Solid histograms refer to the whole–box cluster distributions while dashed histograms refer to the analysis of the mock cluster samples. Left and right panels refer to the sample analysis before and after correcting for Galactic absorbtion mask, respectively. The shaded areas correspond to the 1$\sigma$ interval in the estimate of the $\beta$ parameter for the real Abell/ACO sample.

**Figure 7.** The frequency distributions of the $\Omega_\circ$ values measured by LG–like observers in the $\Lambda$CDM (left panel) and CHDM (right panel) simulations.

# Figure 1

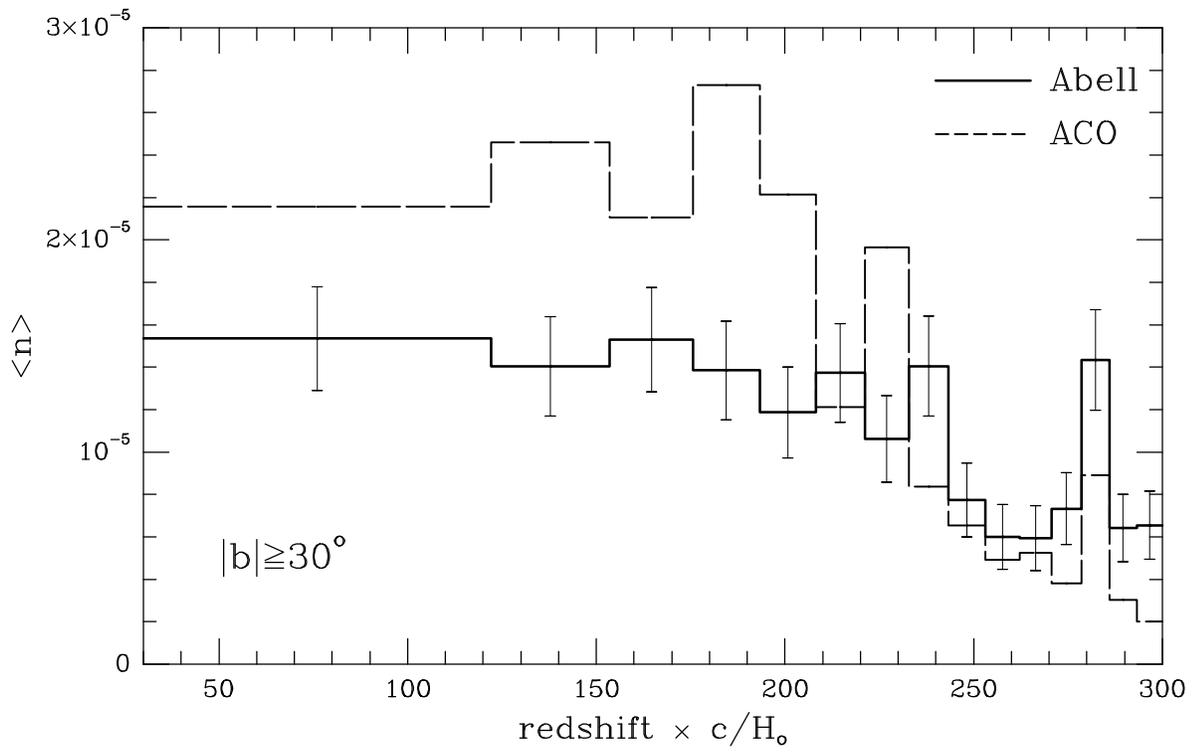

Figure 2

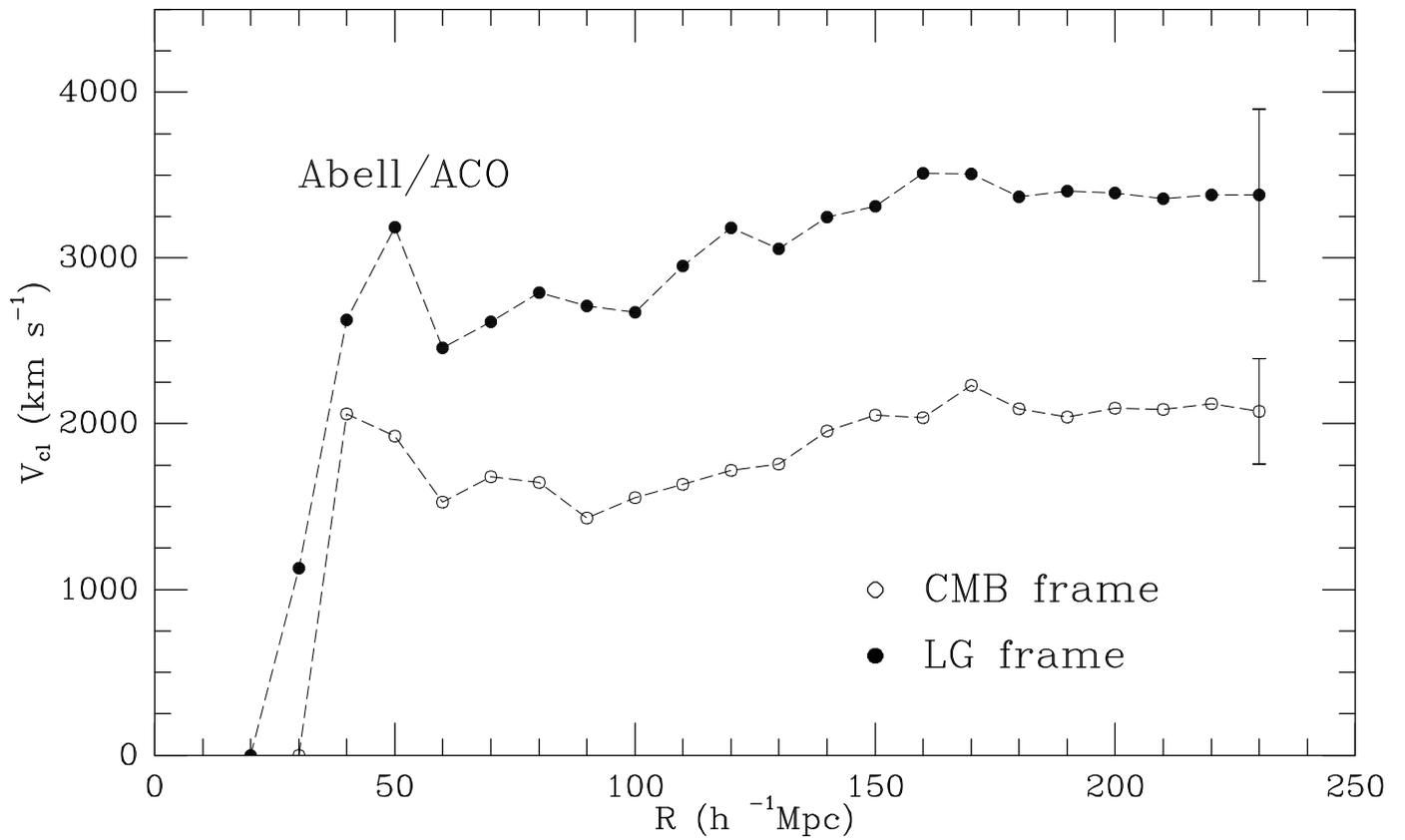

Figure 3

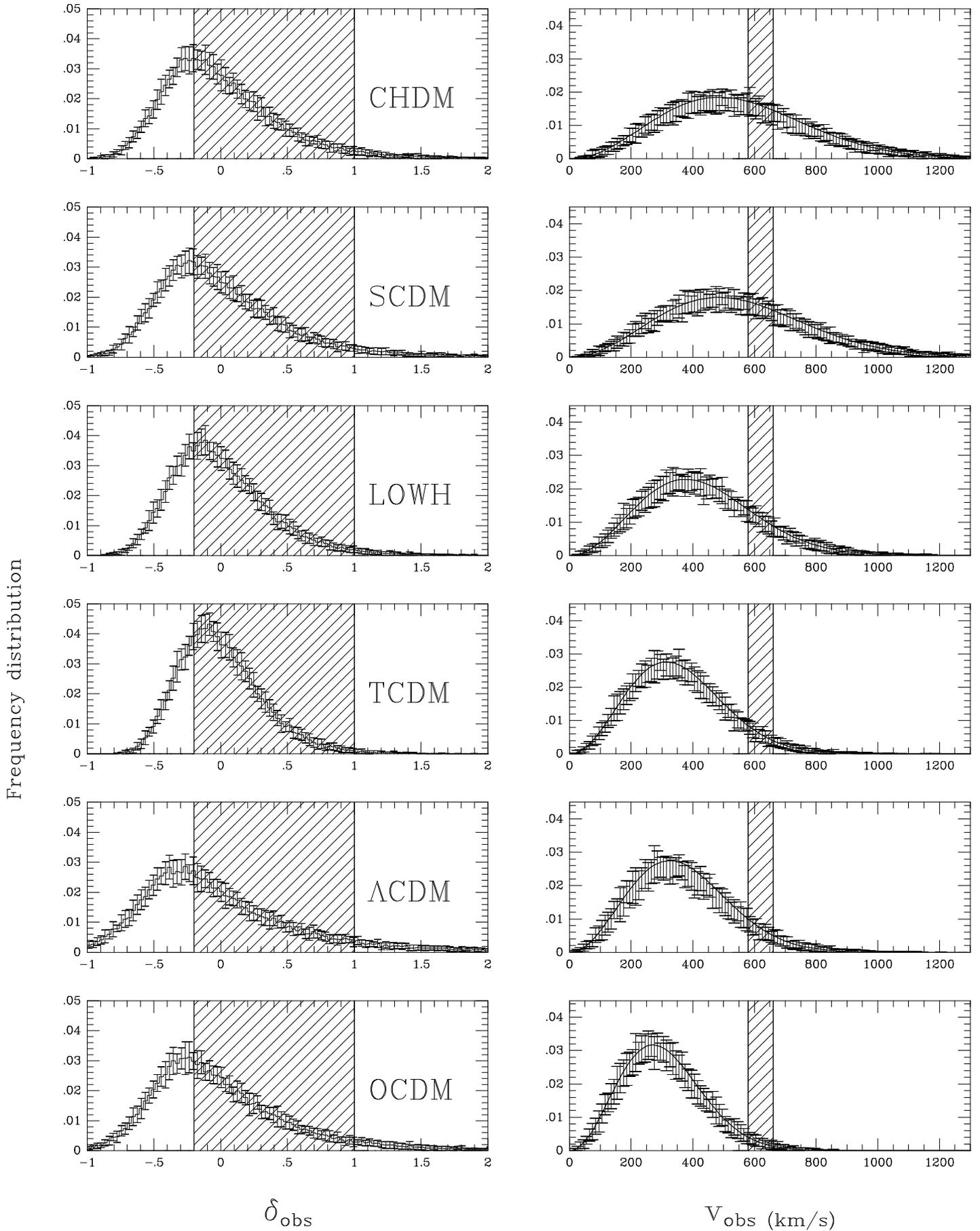

Figure 4

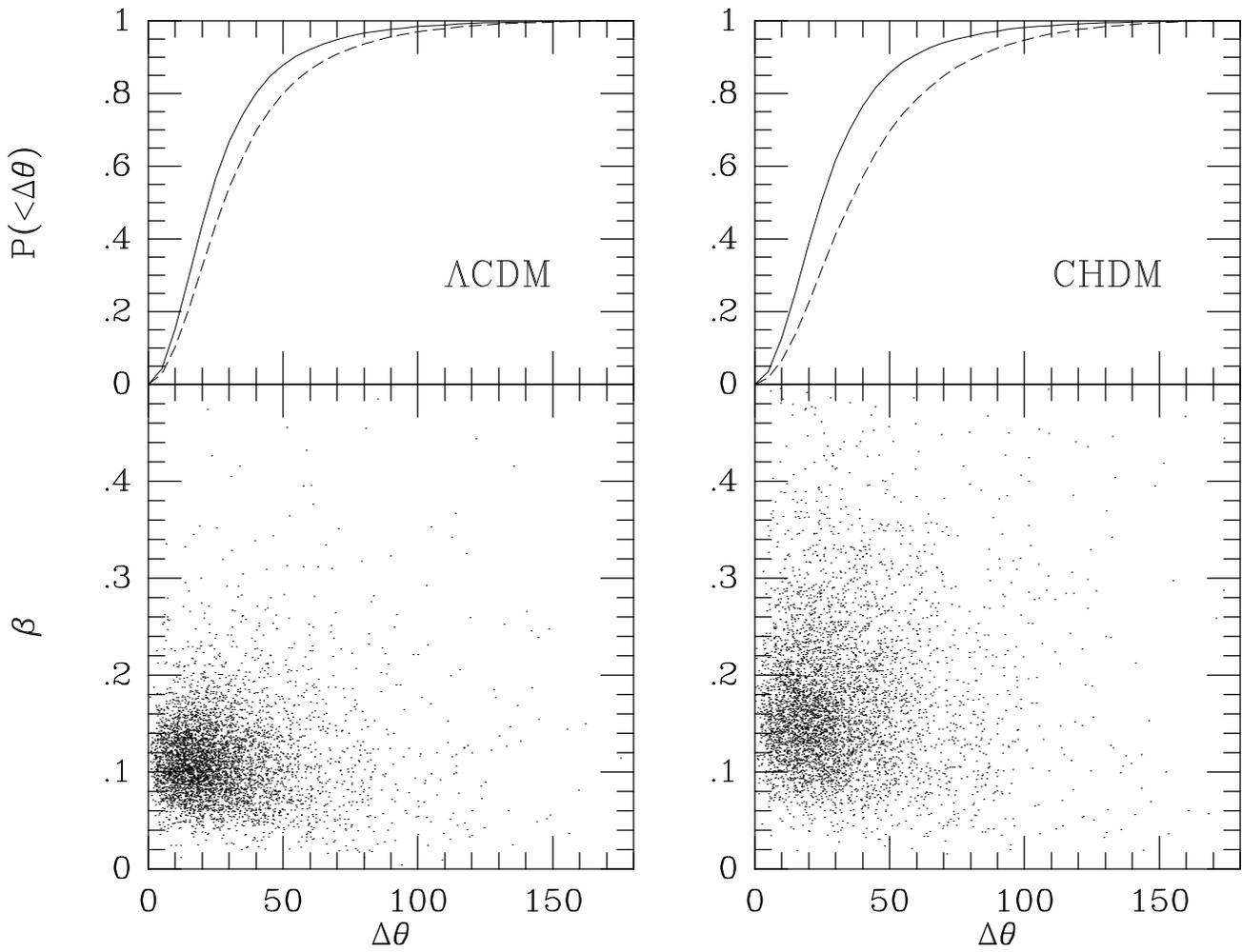

Figure 5

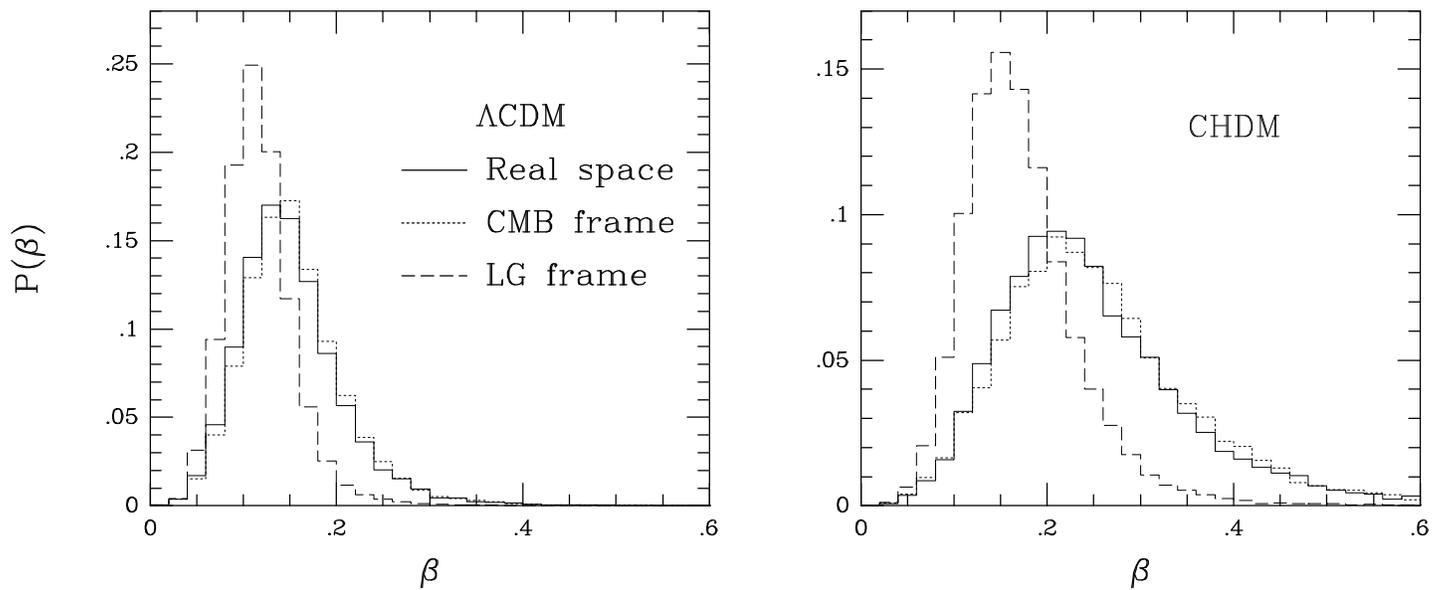

Figure 6

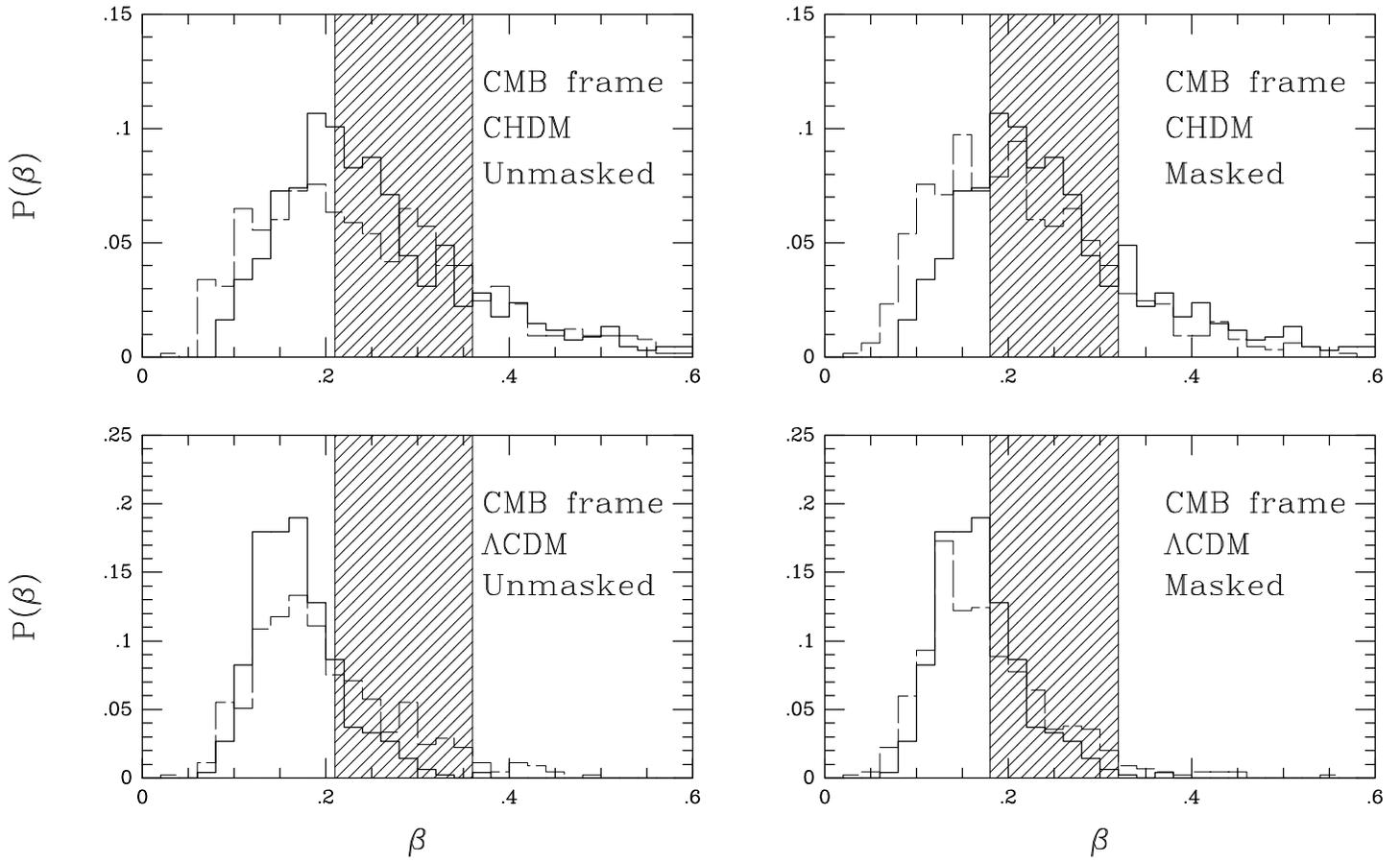

Figure 7

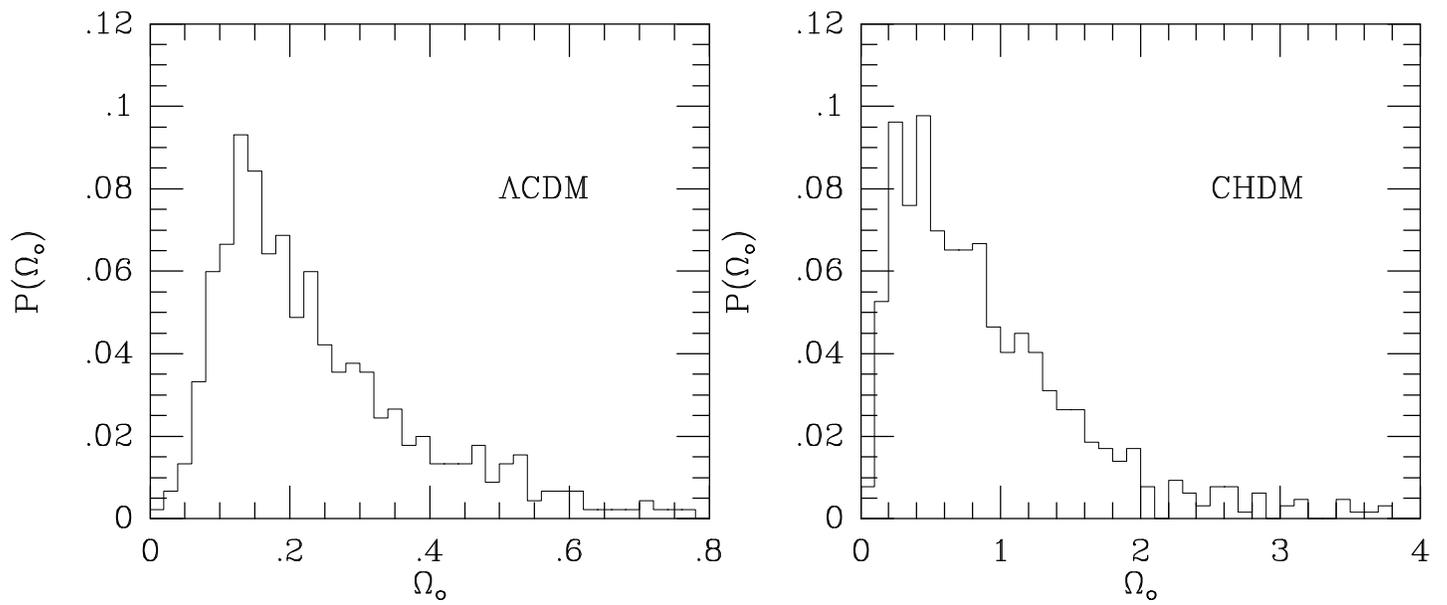